\documentclass[%
 reprint,
 amsmath,amssymb,
 aps,
onecolumn
]{revtex4-1}

\usepackage{subfigure}
\usepackage{graphicx}
\usepackage{dcolumn}
\usepackage{bm}
\usepackage{xcolor}

\begin{document}

\preprint{APS/123-QED}

\title{ Structure of local interactions in complex financial dynamics }

\author{X.F. Jiang}
\affiliation{
 Department of Physics, Zhejiang University, Hangzhou 310027, P.R. China
}\author{T.T. Chen}%
\affiliation{
 Department of Physics, Zhejiang University, Hangzhou 310027, P.R. China
}
\author{B. Zheng}
 \email{zheng@zimp.zju.edu.cn}
\affiliation{
 Department of Physics, Zhejiang University, Hangzhou 310027, P.R. China
}




\date{\today}

\begin{abstract}
With the network methods and random matrix theory, we investigate the interaction structure of
communities in financial markets.
In particular, based on the random matrix decomposition, we clarify that the local interactions between
the business sectors (subsectors) are mainly contained in the sector mode.
In the sector mode,
the average correlation inside the sectors is positive,
while that between the sectors is negative.
Further, we explore the time evolution of the interaction structure of the business sectors, and
observe that the local interaction structure changes dramatically during
a financial bubble or crisis.
\end{abstract}

\maketitle


\section*{Introduction}

Financial markets are dynamic systems with complex interactions,
which share common features in various aspects with those in traditional physics. In recent years, large amounts of historical data have piled up in
stock markets. This allows a quantitative analysis of the financial
dynamics with the concepts and methods in statistical mechanics, and many empirical results have been documented
\cite{man95,gop99,gia01,bou01,bou03,sor03,gab03,qiu06,gar08,she09,she09a,pod09,pod10,pod11,li11,jia12,kum12,pre12,jia13,mun12,gor02,dor00,che13,ou14}.
Very recently, these results are also supported and augmented with the online big data.
For examples, the price change may be predicted with the mood on Twitter \cite{bol11}, and the trading behavior may be quantified with
the Google search volume and Wikipedia topic view times \cite{pre13,moa13}.
Statistical properties of the price fluctuations and cross-correlations
between individual stocks attract much attention, not only scientifically for
unveiling the complex structure and internal interactions of the
financial system, but also practically for the asset allocation
and portfolio risk estimation \cite{man00,bou03}.

In this paper, we are mainly concerned with the cross-correlations between individual stocks in financial markets.
Previous researches have revealed that the cross-correlation between stocks
fluctuates over time \cite{erb94,sol96,le99}.
In particular, it could be enhanced during the recession of business cycles.
Similarly, the international correlation
is stronger in a volatile period.
Based on the time-lag cross-correlations, one
obtains that changes in singular eigenvectors and eigenvalues
are the largest during a financial crisis \cite{pod10}.
Further, individual stocks may form communities due to the cross-correlations between stocks.
A community is also called a business sector, since its stocks usually share common economic properties.
For example, the business sectors have been investigated with the random matrix theory (RMT), in the mature markets such as the American stock market
and Korean stock market \cite{lal99,ple99,ple02a,uts04,qiu10,oh11}, and also in the
emerging markets such as the Chinese stock market and Indian stock market \cite{pan07a,she09,jia12}.
In the emerging stock markets, both standard and unusual business sectors are detected.
The RMT theory may identify the business sectors in a financial market,
but it could not draw the interactions between
the business sectors. Especially, the global price movement in a stock market is usually much stronger than the local one.
To uncover the local interactions between the business sectors, therefore, it is necessary to develop a proper method
to subtract the global price movement.
On the other hand, the structure of financial networks has been gaining
increasing interests \cite{man99,sch09,ken12}. For example, the correlation structure of individual stocks
has been investigated with the minimal spanning tree (MST) and planar
maximally filtered graph (PMFG) \cite{tum05,ast10,poz08,mat10,ken10,kum12}.

The main motivation of this paper is
to investigate the interaction structure between the business sectors in complex financial systems.
Especially, we extract the local interactions between the business sectors with the random matrix decomposition.
The methodology is to combine the RMT theory with various methods
in complex networks, such as the MST tree and PMFG graph, and the theoretical information method (Infomap).
In a complex network, the community structure is the gathering of
nodes into communities with a very high density of internal edges,
including a relatively high density of edges between related communities.
Exploring the community structure with the concepts and methods in statistical mechanics is a very important
topic in the field of complex networks \cite{ros08,new04,new06,li12}.
Our study in this paper will provide a deeper understanding on the communities, i.e., the business sectors,
the interactions between the business sectors, and how this business-sector structure emerges
from the correlations between individual stocks.
In particular, with the random matrix decomposition based on the RMT theory,
we clarify that the local interactions between the business
sectors are mainly contained in the sector mode.
In the sector mode,
the average correlation inside the sectors is positive,
while that between the sectors is negative.
Further, we study the time evolution of the interaction structure.
We discover that the interaction structure of the sector mode varies dramatically with time, while that of the full cross-correlation matrix does not.
Particularly, the interaction structure of the business sectors in the sector mode is simple during a financial bubble or crisis,
and it is dominated by specific business sectors, which are associated with the corresponding economic situation.


\section*{Results}

\subsection*{Interaction structure of business sectors}

Recently a number of activities have been devoted to the study of
the statistical and topological properties of the financial networks \cite{tum05,ast10,poz08,mat10,ken10,kum12}.
In this paper
we concentrate our attention on the interactions between the communities.
Two financial networks, i.e., the MST tree and PMFG graph are first generated from the
cross-correlation matrix of the financial market.
Then the Infomap method is
applied to capture the interaction structure of the communities from the MST tree and PMFG graph.
In common terminology, a community in a financial market is also called a business sector, since its stocks usually share common economic properties.
In the RMT theory, a business sector may split into the positive and negative subsectors,
which are anti-correlated each other \cite{jia12}.
In the present paper, such a phenomenon is also observed in the community structure of the sector mode obtained from the MST tree and PMFG graph.
In these cases, a subsector is usually considered to be a single community,
but for convenience, we may still use the word ``a business sector'' to represent ``a business sector (subsector)''.

To investigate the cross-correlations between individual stocks and the community structure in a stock market,
the data set should contain as many stocks as we can obtain.
On the other hand, the total time length of the time series of stock prices should also be as long as possible.
With these considerations, we obtain the daily data of $259$ weighted stocks in the Shanghai Stock Exchange (SSE),
from Jan., 1997 to Nov., 2007, with 2633 data points in total.
For comparison, we select also $259$ weighted stocks in the New York Stock Exchange (NYSE),
and collect the daily data from Jan., 1990 to Dec., 2006, with 4286 data points in total.
In the selection of these $259$ stocks, we keep the diversity cross different business sectors.
In addition, to further analyze the interaction structure of the communities during a financial crisis,
we collect another data set, with the daily data of $249$ weighted stocks from Oct., 2007 to Nov., 2008,
for both the NYSE and SSE markets. Over ninety-five percent of these $249$ stocks overlap
with the above $259$ stocks.
All these data are obtained from Yahoo Finance (http://finance.yahoo.com).
If the price of a stock in a particular day is absent, it is assumed that the price
is the same as the preceding day \cite{wil07}.
It has been pointed out that the missing data do not lead to artifacts \cite{pan07a}.

\begin{figure}[h]
\centering
\includegraphics[scale=0.45]{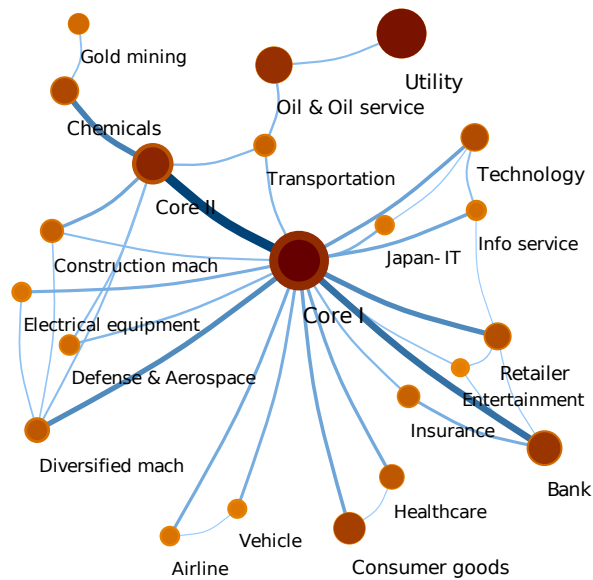}
\caption{\label{fig4} The interaction structure of the business sectors of the PMFG graph for the NYSE market,
plotted with an online tool of MapEquation (http://www.mapequation.org). The size of a node represents the number of stocks inside a community,
and the width of an edge indicates the strength of the interaction between two communities.}
\end{figure}

In Fig.~\ref{fig4}, the community structure of the PMFG graph for the
NYSE market is shown.
That of the MST tree looks similar but
without cycles. Most of the communities
of both the MST tree and PMFG graph can be
identified as the business sectors detected
with the RMT theory \cite{she09,jia12}. In the NYSE market, the business sectors are usually standard.
In the SSE market, however, there exist a few unusual business sectors, whose effects are even stronger
than the standard ones \cite{she09,jia12}. For example, a company
will be specially treated if its financial situation is abnormal.
In this case, the acronym ``ST'' will then be added to the stock name as a prefix.
The abnormal financial situation includes: the
audited profits are negative in two successive accounting
years, and the audited net worth per share is less than the
par value of the stock in the recent accounting year etc.
The acronym ``ST'' will be removed when the financial
situation becomes normal. The so-called ST sector consists of the ``ST'' stocks \cite{she09}.

In Table~\ref{table1}, the business sectors of
the PMFG graph are compared with those from the RMT theory.
In fact, all business sectors (subsectors) identified with the RMT theory can be observed in the communities
of the PMFG graph, but it is not true vice versa. A typical example in the NYSE market is, that four
communities of industrial goods, including Diversity machinery, Construction
machinery, Defense \& Aerospace, and Electrical equipment, can not be found in
the RMT sectors.
On the other hand, the stocks in different RMT sectors may overlap each other, but this is not
allowed in the communities of the PMFG graph.
In the RMT theory, for examples, both $\lambda_{6}^{-}$ and $\lambda_{9}^{+}$ represent the
finance business sectors, which share most common stocks.

To quantitatively characterize the interaction structure of the business sectors,
we define the average correlation inside a business sector as
\begin{equation}
\overline{C}_{ij}^{in}=\frac{1}{E}\sum_{p}\sum_{i\not=j}C_{ij}(p),\label{eq:4}
\end{equation}
where $C_{ij}(p)$ is the cross-correlation $C_{ij}$ of an edge between $i$ and $j$ stocks belonging to
a same sector $p$, and $E$ is the total number of such edges. For comparison,
we define the average correlation between two business sectors as
\begin{equation}
\overline{C}_{ij}^{be}=\frac{1}{E'}\sum_{p\not=q}\sum_{i\not=j}C_{ij}(pq),\label{eq:10}
\end{equation}
where $C_{ij}(pq)$ is the cross-correlation $C_{ij}$ of an edge between $i$ and $j$ stocks belonging to
different business
sectors $p$ and $q$, and $E'$ is the total
number of such edges. The results of $\overline{C}_{ij}^{in}$ and $\overline{C}_{ij}^{be}$ are shown in Table~\ref{table2}. For the NYSE market,
$\overline{C}_{ij}^{in}$ of the RMT sectors (subsectors),
MST tree and PMFG graph,
are $0.33$, $0.49$ and $0.40$ respectively, which are two to three times as large as the
average cross-correlation, $\overline{C}_{ij}=0.16$.
For the SSE market, $\overline{C}_{ij}^{in}$ are $0.40$, $0.48$ and $0.43$ respectively,
which should be compared with the average cross-correlation, $\overline{C}_{ij}=0.37$.
We also note that $\overline{C}_{ij}^{in}$ of the MST tree and PMFG graph are larger
than those of the RMT sectors.
In this sense, the network methods refine the business sectors obtained with the RMT theory.
By the definition of the communities,
the average correlation inside a community, $\overline{C}_{ij}^{in}$, should be stronger
than that between two communities, $\overline{C}_{ij}^{be}$.
Since there are ten to twenty communities, the total number of edges between two different communities is much larger than
the total number of edges inside a community. Therefore, as can be seen in Table~\ref{table2}, $\overline{C}_{ij}^{be}$ is very close to $\overline{C}_{ij}$,
much smaller than $\overline{C}_{ij}^{in}$.


To further understand the interactions between the business sectors in Fig.~\ref{fig4},
we introduce $\overline{C}_{ij}^{li}$ to denote
the average correlation of the edges between two business sectors with a link,
and $\overline{C}_{ij}^{de}$ to represent the one without links.
For comparison, the average correlation between the positive and negative subsectors in a
particular eigenmode of the RMT theory , $\overline{C}_{ij}^{+-}$,
is also computed\cite{jia12}.
All these results are shown in Table~\ref{table2}.
It is clear that the correlation $\overline{C}_{ij}^{de}$ is
weaker than $\overline{C}_{ij}^{li}$.
In fact, $\overline{C}_{ij}^{be}$ is a weighted average
of $\overline{C}_{ij}^{li}$ and $\overline{C}_{ij}^{de}$.
For the RMT theory, $\overline{C}_{ij}^{+-}$ is not only much smaller than $\overline{C}_{ij}^{in}$, but also somewhat
smaller than $\overline{C}_{ij}^{be}$, although both the positive and negative subsectors
formally belong to a same sector.
This is a direct indication of the anti-correlation between the positive and negative subsectors
in a particular eigenmode \cite{jia12}.

To the best of our knowledge, this is the first time that one reveals the interaction
structure of the business sectors in the financial markets.
In a certain sense, the community structure of the MST tree or PMFG graph
may provide a more comprehensive understanding of
the interactions between the business sectors, compared to the RMT theory.
However, we observe that in the community structure of the PMFG graph in Fig.~\ref{fig4}, there are two dominating communities, i.e., the so-called Core I and Core II,
which can not be identified as the standard business
sectors. Careful examination shows that the stocks in Core I and Core II are mainly those
with relatively large components in the
eigenvector of the largest eigenvalue in the RMT theory.
In fact, those stocks may be identified as the controlling set of
the financial network with a spectral centrality method, and
Core I and Core II are the network drivers, which
control the pace of the entire system \cite{gal13,nic12,bat12,nep12,ste11}.
To some extent, Core I and Core II form a kind of centers in the community structure,
connecting to the majority of the business sectors.
In contrast, there are only a few links between
other standard business sectors.
In other words, the interaction structure of the business sectors in Fig.~\ref{fig4} is dominated by
Core I and Core II, but it does not represent the real
local interactions between the business sectors.
Therefore, it is necessary to exclude the global price movement, and to extract
the real local interactions between the business sectors.

\subsection*{Local interaction structure in sector mode}

\begin{figure}[h]
\centering
\includegraphics[scale=0.3]{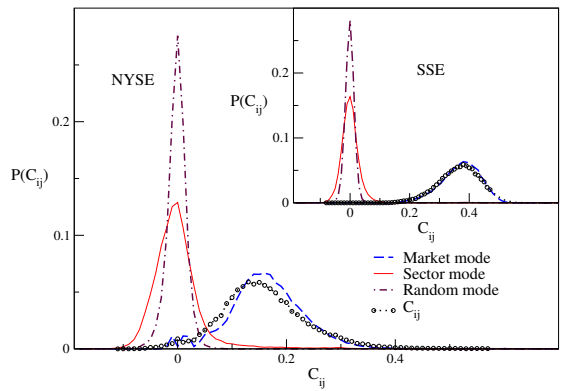}
\caption{\label{fig8} Probability distributions of the matric elements $C_{ij}$ and $C_{ij}^{mode}$ for the NYSE and SSE markets.}
\end{figure}

According to the random matrix decomposition described by Eq.~(\ref{eq:5}) in Sec. Methods,
we may decompose the cross-correlation matrix into $N$ eigenmodes.
With Eq.~(\ref{eq:6}), these eigenmodes are then grouped into
three important modes, which are called the market mode, sector mode
and random mode in this paper. In Fig.~\ref{fig8}, the probability distributions of the matrix elements of the three
modes are shown.
The probability distribution of the matrix elements of the market mode is similar to that
of the full cross-correlation matrix, while
the central peaks of the sector mode and random mode are close to zero.

\begin{figure}[h]
\centering
\includegraphics[scale=0.3]{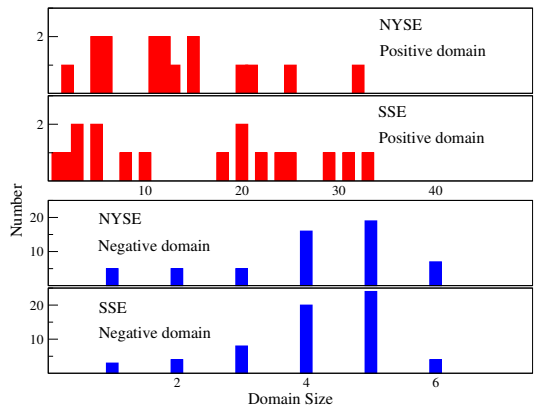}
\caption{\label{fig9} Probability distributions of the domain size in the sector mode for the NYSE and SSE markets.}
\end{figure}

Now we define a domain as a collection of stocks,
in which all cross-correlations between the stocks
are with a same sign, i.e., positive or negative.
The domain formed by positive or negative correlations is called the positive or negative domain, respectively.
For the sector mode, the probability distributions of the domain size are shown in Fig.~\ref{fig9}.
For the NYSE market, there are $17$ positive domains, the
largest domain size is $48$, and the average size
is about $15$. While there are $61$ negative domains, the largest domain size is $7$,
and the average size is about $4$.
The probability distribution of the domain size of the SSE market is
similar to that of the NYSE market.
If a particular stock connects to other two stocks with negative correlations,
there is a low possibility that the correlation between those two domains is also negative.
Thus, it is natural that the average size of the negative domains is small.
The probability distribution of the domain size of both the positive and
negative domains for the random mode is similar to that of the negative domains for the sector mode.
These results imply that the community structure should be mainly
contained in the positive domains of the sector mode.

In fact, the positive domains of the sector mode can be
associated with the business sectors, which are almost identical to
those detected with the RMT theory shown in Table~\ref{table1}.
On the other hand, one can not extract the business sectors from
the negative domains of the sector mode, as well as both the positive
and negative domains of the random mode.
For the market mode or the full cross-correlation matrix, there are very few domains since almost all the correlations are positive,
and these domains are also irrelevant to the business sectors.

\begin{figure}[h]
\centering
\includegraphics[scale=0.35]{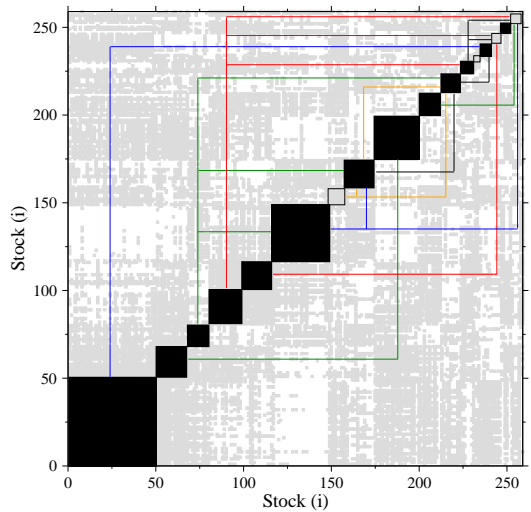}
\caption{\label{fig12} The domain structure for the NYSE market. Gray and white dots denote positive and negative correlations between two stocks, respectively.
The squares along the diagonal line are the domains.
The solid black squares represent the domains which could be associated with the business sectors, while others could not be identified as the standard business sectors. The business sectors from left corner to right are: Industrial goods, Consumer goods, Oil I, Retailer, Finance, Utility, Oil II, IT, Food \& Personal care, Metal Mining, Healthcare, Machinery and Transportation. Here Oil I and Oil II are two groups of Oil companies.}
\end{figure}

To clearly show the domains of the sector mode, and to investigate their interactions,
we plot the interactions between the individual stocks in Fig.~\ref{fig12}.
Gray and white dots denote positive and negative correlations between two stocks, respectively.
The squares along the diagonal line are the domains.
To uncover the interaction structure of the domains,
we calculate the average correlation $\overline{C}_{sec}^{be}\left(pq\right)$ between two
different domains $p$ and $q$.
In Fig.~\ref{fig12}, two domains will be linked with a line,
if $\overline{C}_{sec}^{be}\left(pq\right)$ between this pair of domains
is larger than the average value
of $\overline{C}_{sec}^{be}(pq)$ for all different pairs of domains.

\begin{figure}[h]
\centering
\includegraphics[scale=0.6]{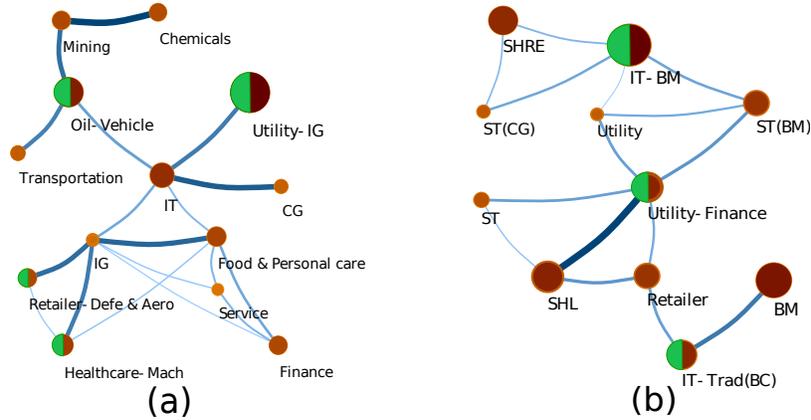}
\caption{\label{fig15} The interaction structure of the business sectors in the sector mode,
obtained from the PMFG graph.
Two semi-circles in a full circle represent a cluster pair,
and the label A-B denotes that the business sectors A and B form a cluster pair.
(a) and (b) display the results for the NYSE and SSE markets respectively. The abbreviations are as follows.
Defe \& Aero: Defense and Aerospace, and Trad(BC): traditional industry of
the Blue-chips stocks.}
\end{figure}

To further understand the business sectors and their local interactions in the sector mode,
we follow the procedure in the previous section, to
generate the PMFG graph from the absolute values of
the matrix elements of the sector-mode cross-correlation matrix $C_{sec}$ defined in Eq.~(\ref{eq:6}), and to extract the community structure
from the PMFG graph with the Infomap method.
The communities of the PMFG graph of the sector mode can also be identified as
the business sectors, which are almost the same as those from the full cross-correlation matrix $C$.
As shown in Fig.~\ref{fig15}, however, the interaction structure of the
business sectors of the sector mode is very different from that of the full cross-correlation matrix.
Core I and Core II in Fig.~\ref{fig4} do not show up anymore.
In contrast, IT, IG, and Food \& Personal care sectors occupy the
central positions for the NYSE market, and IT-BM
and Utility-Finance sectors are the hubs for the SSE market.
In other words, these business sectors actually play a
key role in the interaction structures of the business sectors of the sector mode.
Here, we observe that the interaction structure between the business sectors
in Fig.~\ref{fig15} is similar to that in Fig.~\ref{fig12}.

To briefly summarize, with the random matrix decomposition, one may
remove the global price movement and background noisy correlations described by the market mode and random mode respectively, and identify
the local interaction structure of the business sectors in the sector mode.
The sector mode which reflects the local interactions of the business sectors, plays an important role in the financial market.
On the other hand, we should note that one can not obtain meaningful business sectors
from the market mode and random mode.

Additionally, looking carefully at the internal interactions in a community,
one may find that a community may split into two subcommunities,
which are connected by the negative correlations.
These two subcommunities are called a cluster pair, which is somewhat similar to the pair of positive and negative subsector
in the RMT theory \cite{jia12}.
In Fig.~\ref{fig15}, two semi-circles with
different colors adhered in a full circle represent a cluster pair.
Four cluster pairs are identified for the NYSE market, including
Defe. \& Aero-Retailer, Oil-Vehicle, Utility-IG and Healthcare-Machinery.
Three cluster pairs are detected for the SSE market, i.e., IT-BM,
Utility-Finance and IT-Trad(BC). The cluster pair, which can not be observed
in the community structure of the full cross-correlation matrix,
is a new finding in the financial systems.

To support the
procedure of taking the absolute values of the negative elements of the sector-mode cross-correlation matrix $C_{sec}$,
we set those negative elements to zeros or small positive values.
The PMFG graph is then
generated from the modified cross-correlation matrix
of the sector mode,
and the community structure is extracted.
The resulting business sectors are similar to
those in Fig.~\ref{fig15}, except for the absence of the cluster pairs, due to the removing of the negative correlations.
These results support the
procedure of taking the absolute values of the negative matrix elements.

To quantify the local interactions of the business sectors in the sector mode,
the average correlations inside a business sector and between two business sectors
are computed according to Eqs.~(\ref{eq:4}) and~(\ref{eq:10}), for the cross-correlation matrices
$C_{sec}$ and $|C_{sec}|$, respectively.
The matrix elements of $|C_{sec}|$ are just the absolute values of those of $C_{sec}$ defined in Eq.~(\ref{eq:6}).
The results are shown in Table~\ref{table3}.
Since the market mode has been removed, the magnitude of the correlation becomes relatively small.
However, the differences between various average correlations are very clear.
The average correlation inside the business sectors
is stronger than that between the business sectors.
Meanwhile, the average correlation between the business sectors with a link is stronger than that without a link.
It should be noted that the correlation between the positive and negative subsectors in a cluster pair
is negative, and this also results in that
the average correlation inside the business sectors for $|C_{sec}|$ is larger than that for $C_{sec}$.

In particular, an important observation is that for the sector-mode cross-correlation matrix ${C}_{sec}$, the
average correlation inside the business sectors is positive, while that
between the business sectors is negative.
Considering that the average correlation for $C_{sec}$ is zero, the negative average
correlation between the business sectors indicates
that the business sectors are anti-correlated each other in the sector mode,
and this is a somewhat surprising result.
This kind of anti-correlations can not be directly
detected in the full cross-correlation matrix, due to the dominating effect of the market mode.
This should be potentially important in both theoretical and practical senses.

\subsection*{Time evolution of interaction structure}

The time evolution of the interaction structure of the business sectors
is an important topic in financial dynamics. Practically, it is also useful for
investors to diversify risk in stock markets. Now, we investigate
how the interaction structure evolves with time, combining the complex network methods
with the random matrix decomposition.
For the NYSE stock market, the time period of the daily data of the $259$ stocks is equally divided into four time windows,
i.e., from Jan., 1990 to Mar., 1994, from Apr., 1994 to Jun., 1998, from Jul., 1998 to Sep., 2002,
and from Oct., 2002 to Dec., 2006, which are denoted by window (a), (b), (c) and (d),
respectively. In the NYSE market, there is a typical financial bubble,
i.e., the Dotcom bubble, which is mainly located in the window (c).
The Dotcom bubble is also referred to as the Internet bubble or information technology bubble.

\begin{figure}[h]
\centering
\includegraphics[scale=0.35]{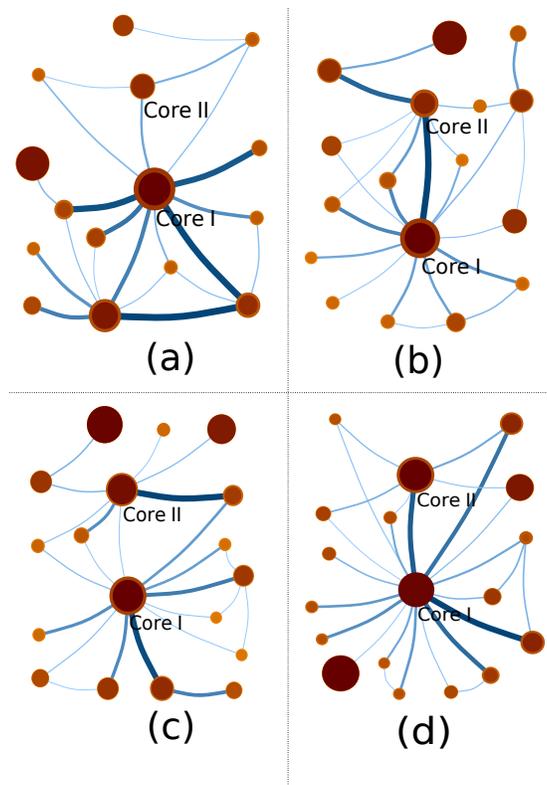}
\caption{\label{fig20} The interaction structure of the business sectors for the full cross-correlation matrix is obtained with the PMFG graph for the NYSE  market in different time windows. The business sectors in different time windows consist of similar stocks.
Window (a): Jan.,1990 to Mar., 1994; window (b): Apr., 1994 to Jun., 1998; window (c): Jul., 1998 to Sep., 2002; window (d): Oct., 2002 to Dec., 2006.
The Dotcom bubble is mainly located in the window (c).}
\end{figure}

In Fig.~\ref{fig20}, the interaction structure of the
full cross-correlation matrix generated by the PMFG graph is shown for the four time windows.
Clearly, the interaction structure for different time windows shares common features.
For example, there are two large cores for all the
four time windows, and most of other
communities connect to the cores. In other words, the interaction structure
of the business sectors is dominated by the cores.
These cores are mainly induced by the market mode, almost time-independent.

\begin{figure}[h]
\centering
\includegraphics[scale=0.4]{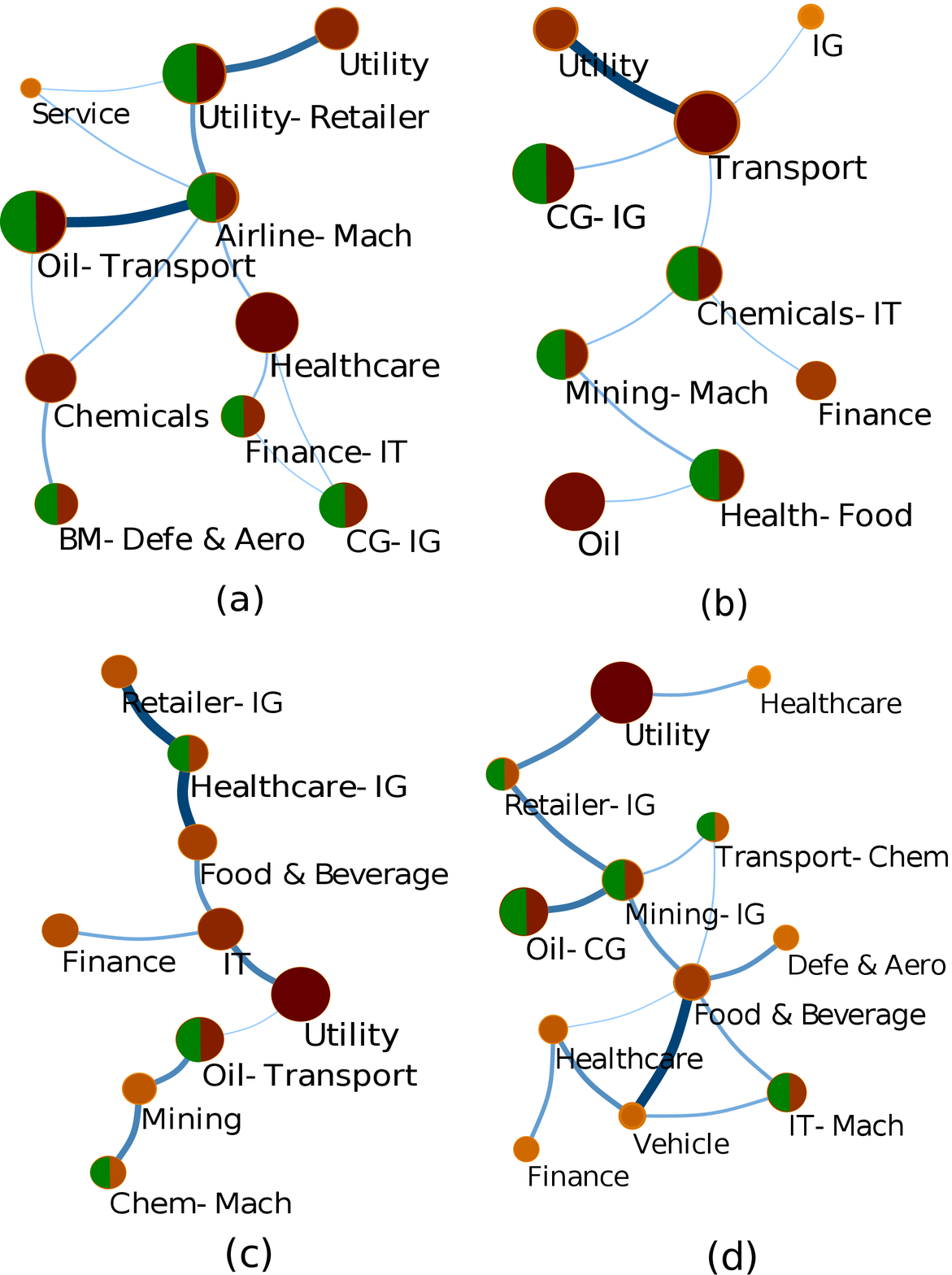}
\caption{\label{fig21} The interaction structure of the business sectors in the sector mode is obtained with the PMFG graph for the NYSE market in different time windows. The time windows are the same as in Fig.~\ref{fig20}.}
\end{figure}

For comparison, the local interaction structure of the business sectors in the sector mode is also extracted
and displayed for the four time windows in Fig.~\ref{fig21}.
Obviously, it varies with time, in contrast to that obtained from the full cross-correlation matrix.
This time-dependent interaction structure is in spirit consistent with the result in Ref.~\cite{ast10}, in which
the network structure of individual stocks is
very different before, during and after a financial crisis.
More specifically, the interaction structure of
the window (c) is very simple, compared to others. The IT sector plays an important
role in the window (c), while it is not so important in other
windows. We believe that
the simple interaction structure is driven by the Dotcom bubble in that period of time.
This IT-dominating structure is reasonable, since the IT industry is very booming and becomes the
core of the market during the Dotcom bubble.  

\begin{figure}[h]
\centering
\includegraphics[scale=0.53]{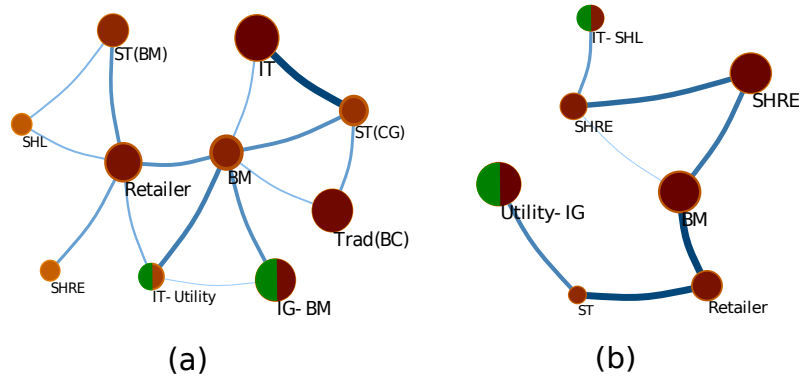}
\caption{\label{fig22} The interaction structure of the business sectors in the sector mode is obtained with the PMFG graph for the SSE market in two time windows. Window (a): Jan., 1997 to Mar., 2002; window (b): Apr., 2002 to Nov., 2007. The huge bubble induced by the real estate industry is contained in the window (b).}
\end{figure}

A similar result is also obtained for the SSE stock market.
We equally divide the time series of the daily data of the $259$ stocks into two time windows. The window (a) is a standard
period, from Jan., 1997 to Mar., 2002, and the window (b) is from Apr., 2002 to Nov., 2007, which contains a huge financial bubble from year
2003 to 2007.
The local interaction structure of the business sectors in the sector mode extracted for the two time
windows is displayed in Fig.~\ref{fig22}. Clearly, the interaction structure
of the window (b) is simple,
and the SHRE (Shanghai Real Estate) sectors including the Basic Material sector are dominating.
Actually, the huge bubble in the window (b) is mainly induced by
the booming of the real estate industry during that time period.

\begin{figure}[h]
\centering
\includegraphics[scale=0.7]{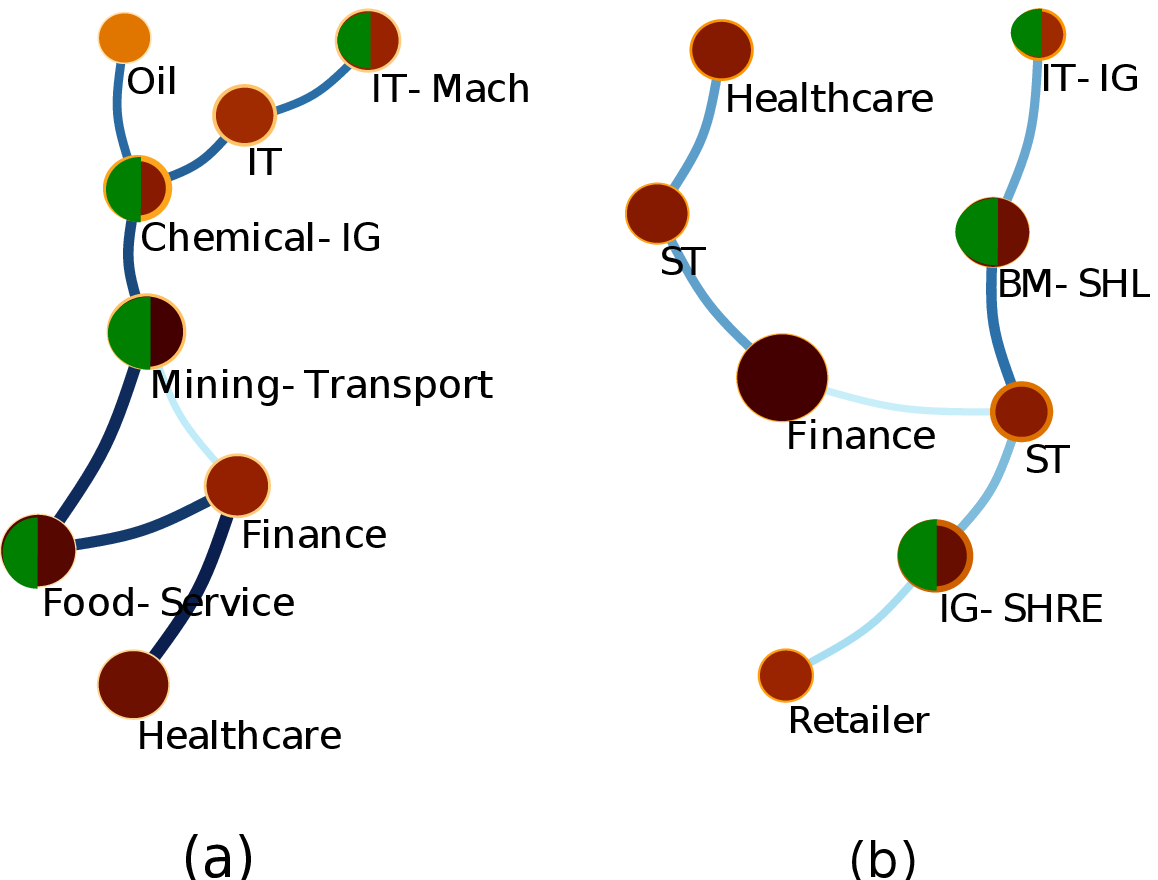}
\caption{\label{fig23} The interaction structure of the business sectors in the sector mode is obtained with the PMFG graph in the time period
of the subprime mortgage crisis. (a): for the NYSE market, from Oct., 2007 to Nov, 2008; (b): for the SSE market, from Oct., 2007 to
Nov., 2008.
}
\end{figure}

To further support our results, we analyze another data set, which contains
the daily data of the $249$ stocks from Oct., 2007 to Nov., 2008, for both the NYSE and SSE markets.
During this time period, both stock markets experienced the subprime mortgage crisis.
The local interaction structure of the business sectors in the sector mode is shown in Fig.~\ref{fig23}.
A relatively simple structure is again observed for both stock markets.
Since the subprime mortgage crisis is induced from finance and banking, the Finance sector
is important in the interaction structure of the business sectors, especially in the NYSE market. Further, the crisis leads to an economic recession,
and the ST sector becomes dominating for the SSE market.

Briefly speaking, the interaction structure of the business
sectors in the sector mode changes dramatically in different time
periods. Especially, it is rather simple during a financial bubble or crisis, dominated by specific business sectors,
which are associated with the corresponding economic situation.

\begin{figure}[h]
\centering
\includegraphics[scale=0.8]{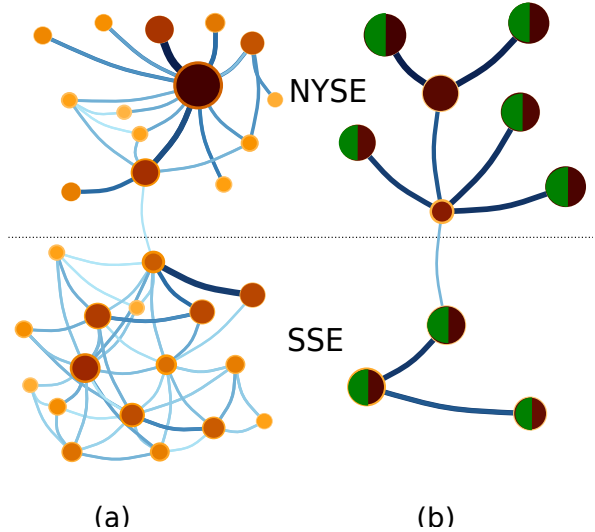}
\caption{\label{fig24} The interaction structure of the business sectors
is obtained with the PMFG graph for
 a combination of the NYSE and SSE markets. (a) and (b) are for the full cross-correlation
 matrix and the sector mode respectively. The business sectors of two
 markets are separated with the dashed line.}
\end{figure}

Finally, we investigate whether there exist interactions between the NYSE and SSE markets at the level of the business sectors.
For this purpose, we combine the NYSE and SSE markets to form a single market.
The overlapping time period of the data sets for the two markets is from Jan., 1997 to Dec. 2006.
We repeat the analysis as described in the previous texts.
In Fig.~\ref{fig24}, the interaction structure of the business sectors is displayed.
Although some stocks of the SSE market may mix into the business sectors of the NYSE market and vice versa,
no interactions at the level of the business sectors are detected between the NYSE and SSE markets,
for either the full cross-correlation matrix or the sector mode.
The only link connecting the two markets is from the method itself, i.e., the graph should not be disconnected.
However, one observes that the number of the business sectors in the sector mode is greatly reduced,
and more cluster pairs emerge. This is because the cross-correlations between the NYSE and SSE markets are weak,
and the stocks tend to form larger communities.
We have also performed such an analysis in different time windows, including that with a financial bubble or crisis.
The results remain qualitatively unchanged. In other words, the interactions between the NYSE and SSE markets are
not so strong in the past years, at least in the level of the business sectors.

\section*{Discussion}

In this paper, we investigate the communities and their interactions in complex financial systems,
with the concepts and methods in statistical mechanics.
We generate the interaction networks from
the cross-correlation matrix of the financial market, based on the MST tree and PMFG graph.
With the Infomap method, we extract the communities and their interaction structure.
In the financial markets, the communities could be associated with the business sectors (subsectors).
Thus the economic implication of the interaction structure of the communities is revealed.

Based on the random matrix decomposition, the market mode and random mode
of the price movement can be subtracted. Thus we demonstrate that the
local interactions of the business sectors
are mainly contained in the sector mode, from the perspective of both the domain
structure and the PMFG graph.
In the sector mode,
the average correlation inside a business sector is positive,
while that between two business sectors is negative. In other words,
the business sectors are anti-correlated each other.
Meanwhile, a cluster pair structure is also observed.
The interaction structure of the business sectors in the sector mode
varies dramatically with time.
In particular, it becomes simple during a financial bubble or crisis, dominated by specific business sectors.
The interaction structure of the business sectors
of the full cross-correlation matrix does not change so much with time.

\section*{Methods}

\subsection*{Random matrix decomposition}
The logarithmic price return of the \textit{$i$}-th stock
over a time interval $\Delta t$  is defined by
\begin{equation}
R_{i}\left(t\right)\equiv ln\, P_{i}\left(t+\Delta
t\right)-ln\, P_{i}\left(t\right),\label{eq:1}
\end{equation}
where $P_{i}\left(t\right)$ represents the stock price at
time $t$. For the daily data, we typically take $\Delta t=1$. To ensure different stocks with an equal weight, we
introduce the normalized price return
\begin{equation}
r_i\left(t\right)=\frac{R_{i}(t)-\langle R_{i}(t)\rangle}{\sigma_{i}},\label{eq:2}\end{equation}
where $\langle \cdots\rangle$ is the average over time $t$, and
$\sigma_{i}=\sqrt{\langle R_{i}^{2}\rangle-\langle R_{i}\rangle^{2}}$ denotes the standard
deviation of $R_{i}$. Then, the matrix elements of the
cross-correlation matrix $C$ are defined by the equal-time correlations
\begin{equation}
C_{ij}\equiv\langle r_{i}(t)r_{j}(t)\rangle.\label{eq:3}
\end{equation}
By the definition, $C$ is a real symmetric matrix with $C_{ii}=1$,
and $C_{ij}$ is valued in the range $\left[-1,1\right]$.

Statistical properties of the eigenvalues of the so-called Wishart matrix
are known, and the Wishart matrix is derived from uncorrelated time series
with a finite length. We denote the total number of stocks by $N$, and the total time length of the data by $T$. In the limit $N\rightarrow\infty$ and $T\rightarrow\infty$ with $Q\equiv T/N\geq1$,
the probability distribution of the eigenvalues $\lambda$ is given
by \cite{dys71,sen99}
\begin{equation}
P_{rm}\left(\lambda\right)=\frac{Q}{2\pi}\frac{\sqrt{\left(\lambda_{max}^{ran}-\lambda\right)\left(\lambda-\lambda_{min}^{ran}\right)}}{\lambda},\end{equation}
with the upper and lower bounds $\lambda_{min\left(max\right)}^{ran}=\left[1\pm\left(1/\sqrt{Q}\right)\right]^{2}$.

Large eigenvalues of the cross-correlation matrix in financial markets deviate from
$P_{rm}\left(\lambda\right)$ of the Wishart matrix. It suggests that there exist non-random interactions between individual stocks,
and the largest eigenvalue represents the global price movement of the market. In our notation, the eigenvalues
are arranged in the order of $\lambda_{\alpha}>\lambda_{\alpha+1}$, with $\alpha =0, \cdots , N$.
The eigenmodes with large eigenvalues can be
associated with business sectors. Taking into account the signs of
the components of the eigenmodes, the sector of $\lambda_{\alpha}$ may be separated into two subsectors, i.e., the positive and negative subsectors \cite{jia12}.

Finally, based on the RMT theory, the cross-correlations $C_{ij}$ can be decomposed into
different eigenmodes,
\begin{equation}
C_{ij}=\sum_{\alpha=1}^{N}\lambda_{\alpha}C_{ij}^{\alpha},\quad C_{ij}^{\alpha}=u_{i}^{\alpha}u_{j}^{\alpha}, \label{eq:5}
\end{equation}
where $\lambda_{\alpha}$ is the $\alpha$-th eigenvalue of $C_{ij}$, $u_{i}^{\alpha}$ is the
$i$-th component in the $\alpha$-th eigenvector. Thus, $C_{ij}^{\alpha}$ represents the cross-correlation
in the $\alpha$-th eigenmode. In order to uncover the interaction structure in different eigenmodes, we
define three mode cross-correlation matrices by
\begin{equation}C_{mode,ij}=\sum^{mode}_{\alpha}C_{ij}^{\alpha}. \label{eq:6}\end{equation}
For the market mode, $C_{mar,ij}=C_{ij}^{0}$, corresponding to the
largest eigenvalue of $C_{ij}$. For the sector mode, $C_{sec,ij}=\sum_{\alpha=1}^{11}C_{ij}^{\alpha}$ for the NYSE market, and
$C_{sec,ij}=\sum_{\alpha=1}^{6}C_{ij}^{\alpha}$ for the SSE market. For the random mode, we simply take $C_{ran,ij}=\sum_{\alpha=50}^{258}C_{ij}^{\alpha}$
for both the NYSE and SSE markets.

\subsection*{Methods of complex networks}
In principle, the matrix elements $C_{ij}$ contain full information of the cross-correlations between stocks,
but carry also strong noises.
To sketch the backbone of the interactions between stocks, the edges should be greatly discarded with
respect to the complete graph. To this aim, the graph should be kept as simple as possible, but
important correlations should remain. The MST tree
is therefore the
simplest connected graph \cite{kru56,pri57}, in which there are no cycles. Moreover, the MST tree may give
the simplest subdominant organization structure of individual stocks in a direct scheme.

We briefly describe the MST method as follows:
$(1)$ Rank the correlations $C_{ij}$ between stocks $i$ and $j$ in a list by descending order.
$(2)$ Capture the first one in the list, and add the nodes $i$ and $j$, and the edge between $i$ and $j$ to the graph.
$(3)$ Take the next one in the list, and add the corresponding edges and nodes to the graph
if it could be linked to the existing nodes, and would not form cycles; otherwise ignore it.
$(4)$ Repeat step (3) until all links have been visited.

Assuming that there are $N$ stocks, the resulting MST tree consists of $N$ nodes and $N-1$ edges, and maximizes the sum of $C_{ij}$ in the tree graph.
The MST tree is an efficient approach to capture the most
relevant correlations of each node. However, it is a strong constraint that only the tree
structure is permitted.
All those edges are rejected when they form cycles with the existing edges, even if they may represent
significant correlations.

To improve the method, the PMFG graph is proposed \cite{tum05}.
The solution generates a graph embedded on a surface with a particular genus $g$. The genus of a surface is the number of holes in the surface, and $g=0$ corresponds to a topological sphere, $g=1$ to a torus, $g=2$ to
a double torus, etc.
The PMFG algorithm is generally the same as the MST one, except for the step $(3)$.
  It is replaced by the condition that the resulting graph should be embeddable on a surface with genus $g$.
  The graph generated by the PMFG algorithm is a triangulation of the surface, and it contains $3N+6g-6$ links, which maximize the sum of $C_{ij}$.
  The simplest case is the graph with $g=0$. In some sense, the PMFG graph could be viewed as the minimum extension towards the complexity after the MST tree.

The next step after the extraction of the MST tree and PMFG graph,
we apply the Infomap method proposed in Ref.~\cite{ros08},
to investigate the community structure of the financial network.
The method is based on a random walk on the network to capture the
information flow, then to extract the community structure from the regularity of the data.
Detailed descriptions of the method are referred to Ref.~\cite{ros08}.
With the Infomap method, we can identify the community
structure for both the MST tree and PMFG graph.

Further, with the mode cross-correlation matrix $C_{mode}$ at hand, we may construct the MST tree and PMFG graph for different modes, and investigate which modes dominate the business sectors and their interactions. This is an important step in the exploration of the interaction structure in financial systems.

\begin{acknowledgments}
This work was supported in part by NNSF of China
under Grant Nos. 11375149 and 11075137, and Zhejiang Provincial
Natural Science Foundation of China under Grant No. Z6090130.
\end{acknowledgments}


\begin{table}
\caption{The PMFG communities and the corresponding RMT subsectors.
$\lambda_{\alpha}^{+}$ and $\lambda_{\alpha}^{-}$ represent the positive and negative subsectors, respectively, in
the $\alpha$-th eigenmode of the RMT theory \cite{jia12}.
The abbreviations are as follows.
Info. service:
Information service, Japan-IT: Japan IT company,
BM: Basic materials, CG: Consumer goods, mach.: machinery,
SHRE: Shanghai Real Estate, SHL: Shanghai Local company, IG: Industrial goods,
ST (BM\&IG): BM and IG with ST, ST (FI\&CG): Finance and
CG with ST.}
\label{table1}
\centering
\begin{tabular}{cc|cc}
\hline
NYSE & & SSE & \tabularnewline
\hline
PMFG & RMT & PMFG & RMT\tabularnewline
\hline
Utility & $\lambda_{1}^{+}$ & Finance & $\lambda_{5}^{-}$ \tabularnewline
Transportation & $\lambda_{11}^{-}$ & IT & $\lambda_{2}^{-}$\tabularnewline
Oil & $\lambda_{2}^{-}$ & SHRE & $\lambda_{3}^{-}$ \tabularnewline
Gold mining & $\lambda_{5}^{-}$ &  ST (BM\&IG) & $\lambda_{1}^{-}$ \tabularnewline

Retailer & $\lambda_{7}^{-}$,$\lambda_{8}^{+}$ &  ST (FI\&CG) & $\lambda_{3}^{-}$ \tabularnewline
Info. service & $\lambda_{1}^{-}$ &  IG & $\lambda_{6}^{+}$ \tabularnewline
Technology & $\lambda_{3}^{+}$,$\lambda_{7}^{+}$ & CG &\tabularnewline
Japan-IT & $\lambda_{10}^{-}$ & Cement & \tabularnewline

Healthcare & $\lambda_{8}^{-}$ & BM \tabularnewline
CG & $\lambda_{3}^{-}$ & SHL & \tabularnewline
Insurance & $\lambda_{6}^{-}$,$\lambda_{9}^{+}$ & Utility\tabularnewline
Bank & $\lambda_{6}^{-}$,$\lambda_{9}^{+}$ & Retailer &  \tabularnewline

Airline & $\lambda_{11}^{-}$ & Healthcare & \tabularnewline
BM & & & \tabularnewline
Diversity mach. & \tabularnewline
Construction mach. & \tabularnewline
Electrical equipment & \tabularnewline
Defense\&Aerospace & \tabularnewline
Vehicle & &  &  \tabularnewline
Entertainment & &  & \tabularnewline

\hline
\end{tabular}
\end{table}

\begin{table}
\caption{The average cross-correlation $\overline C_{ij}$ for the PMFG graph,
MST tree and RMT sectors (subsectors). $\overline{C}_{ij}^{in}$ and $\overline{C}_{ij}^{be}$ represent
the average correlations inside and between the communities respectively. $\overline{C}_{ij}^{li}$
and $\overline{C}_{ij}^{de}$
are the average correlations between two different communities with a link and without
a link.  $\overline{C}_{ij}^{+-}$
is the average correlation between the positive and negative subsectors in an
eigenmode \cite{jia12}.}
\label{table2}
\centering
\begin{tabular}{cc|c|cc|cc|c}
\hline
& & $\overline{C}_{ij}$ & $\overline{C}_{ij}^{in}$ & $\overline{C}_{ij}^{be}$ & $\overline{C}_{ij}^{li}$ & $\overline{C}_{ij}^{de}$ & $\overline{C}_{ij}^{+-}$\tabularnewline
\hline
NYSE & PMFG &$0.16$ & $0.46$ & $0.15$ & $0.18$& $0.13$ &        \tabularnewline
 & MST &$0.16$ & $0.49$ & $0.16$ & $0.19$& $0.12$ &        \tabularnewline
     & RMT &$0.16$ & $0.33$ & $0.16$ &       &        & $0.15$ \tabularnewline
\hline
SSE & PMFG &$0.37$ & $0.43$ & $0.36$ & $0.38$ & $0.32$ &        \tabularnewline
 & MST &$0.37$ & $0.48$ & $0.36$ & $0.39$ & $0.31$ &        \tabularnewline
& RMT &$0.37$ & $0.40$ & $0.34$ &       &        & $0.32$ \tabularnewline
\hline
\end{tabular}
\end{table}

\begin{table}
\caption{The average cross-correlation $\overline{C}_{ij}$ of the sector mode
for the PMFG graph. All values in the table have been multiplied by a factor of $10^{2}$.
$\overline{C}_{ij}^{in}$ and $\overline{C}_{ij}^{be}$
represent the average correlations inside and between
the communities respectively. $\overline{C}_{ij}^{li}$ and $\overline{C}_{ij}^{de}$
are the average correlations
between two different communities with a link and without a link.
$\overline{C}_{ij}^{+-}$ denotes the average correlation
between the positive and negative subsectors in a cluster pair.}
\label{table3}
\centering
\begin{tabular}{cc|c|cc|cc|c}
\hline
& & $\overline{C}_{ij}$ & $\overline{C}_{ij}^{in}$  & $\overline{C}_{ij}^{be}$& $\overline{C}_{ij}^{li}$ & $\overline{C}_{ij}^{de}$ & $\overline{C}_{ij}^{+-}$
\tabularnewline
\hline
NYSE  & $C_{sec}$    & $0$ & $7.4$ & $-0.7$ & $-0.4$ & $-0.8$ & $-0.9$ \tabularnewline
     & $|C_{sec}|$  & $3.2$ & $9.7$  & $2.5$ & $2.8$  & $2.4$ & $2.3$   \tabularnewline
\hline
SSE  & $C_{sec}$    & $0$ & $1.3$ & $-0.2$ & $-0.1$ & $-0.2$ & $-0.2$   \tabularnewline
     & $|C_{sec}|$  & $2.1$ & $3.8$ & $1.9$ & $2.3$  & $1.8$ & $1.7$    \tabularnewline
\hline
\end{tabular}
\end{table}

\section*{\textbf{Author contribution}}
X.F.J. and B.Z. conceived the study. X.F.J., T.T.C and B.Z. designed and performed the research.
X.F.J. and T.T.C performed the statistical
analysis of the data. X.F.J. drafted the manuscript. B.Z. reviewed and approved the manuscript.

\section*{\textbf{Additional information}}
\textbf{Competing financial interests}: The authors declare no competing financial interests.


\end{document}